\documentclass[conference]{IEEEtran}
\usepackage{cite}
\usepackage{amsmath,amssymb,amsfonts}
\usepackage{algorithmic}
\usepackage{graphicx}
\usepackage{subcaption}
\usepackage[labelformat=parens,labelsep=quad,skip=3pt]{caption}
\usepackage{textcomp}
\usepackage{xcolor}
\usepackage{fancyhdr}
\usepackage[hyphens]{url}
\usepackage{booktabs}
\usepackage{array, makecell}

\def\BibTeX{{\rm B\kern-.05em{\sc i\kern-.025em b}\kern-.08em
    T\kern-.1667em\lower.7ex\hbox{E}\kern-.125emX}}
    
\author{
    \IEEEauthorblockN{Kaoutar El Maghraoui\IEEEauthorrefmark{1}, Lorraine M. Herger\IEEEauthorrefmark{1}, Chekuri Choudary\IEEEauthorrefmark{2}, Kim Tran\IEEEauthorrefmark{2}, Todd Deshane\IEEEauthorrefmark{1}, David Hanson\IEEEauthorrefmark{1}}
    \IEEEauthorblockA{\IEEEauthorrefmark{1}IBM T. J. Watson Research Center, New York, USA
    \\\{kelmaghr, herger, kvtran, dhhanson\}@us.ibm.com}
    \IEEEauthorblockA{\IEEEauthorrefmark{2}IBM Systems, Texas, USA
    \\\{Chekuri.Choudary, Todd.Deshane\}@ibm.com}
}

\title{Performance Analysis of Deep Learning Workloads on a Composable System} 
\begin{document}

\maketitle
%\thispagestyle{firstpage}
%\pagestyle{plain}

%%%%%% -- PAPER CONTENT STARTS-- %%%%%%%%

\begin{abstract}
A composable infrastructure is defined as resources, such as compute, storage, accelerators and networking, that are shared in a pool and that can be grouped in various configurations to meet application requirements. This freedom to 'mix and match' resources dynamically allows for experimentation early in the design cycle, prior to the final architectural design or hardware implementation of a system. This design provides flexibility to serve a variety of workloads and provides a dynamic co-design platform that allows experiments and measurements in a controlled manner. For instance, key performance bottlenecks can be revealed early on in the experimentation phase thus avoiding costly and time consuming mistakes. Additionally, various system-level topologies can be evaluated when experimenting with new System on Chip (SoCs) and new accelerator types. This paper details the design of an enterprise composable infrastructure that we have implemented and made available to our partners in the IBM Research AI Hardware Center (AIHC). Our experimental evaluations on the composable system give insights into how the system works and evaluates the impact of various resource aggregations and reconfigurations on representative deep learning benchmarks. 

\end{abstract}

\section{Introduction}
% Lorraine, Kaoutar, Chekuri

\begin{figure}[ht!]
    \centering
    \includegraphics[width=7cm]{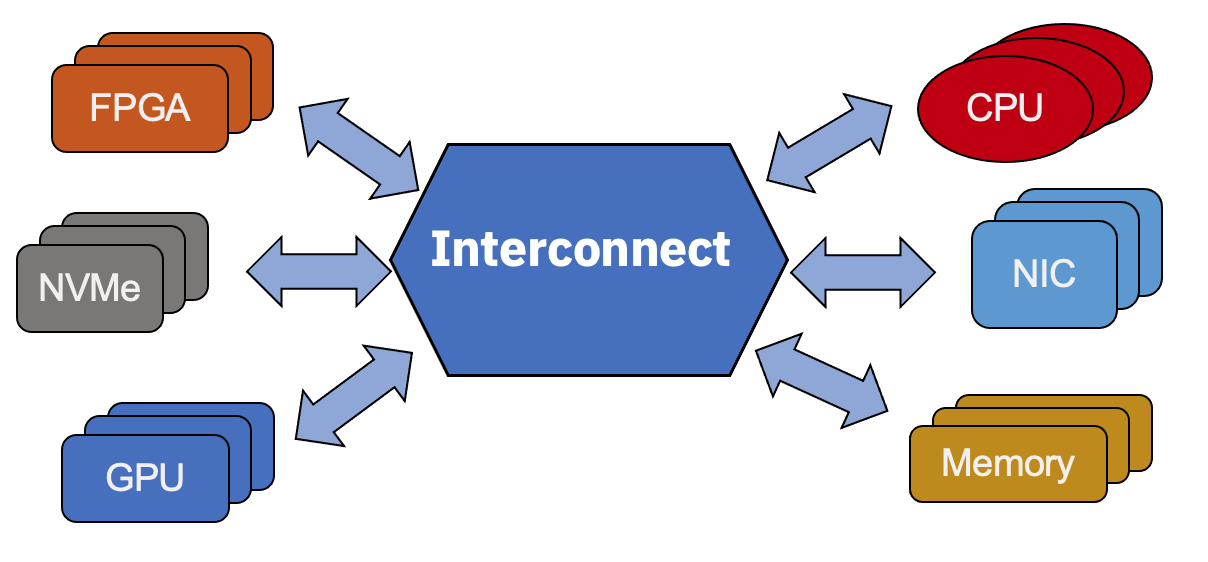}
    \caption{Composable Server Concept}
    \label{fig:comp_concept}
\end{figure}

Today’s data centers are often plagued by resource wasting and under-utilization across a host of workloads. This problem is often due to the prevalent method of delivering resources in fixed ratios, which forces coupling of all the server’s components. While academic projects may not view these issues as compelling, for industrial enterprises prolonged under-utilization can mean failure and bankruptcy. Moreover, resources for a large percentage of workloads are often over-provisioned due to an incomplete understanding of the component-level hardware requirements of the workloads. The composable server architecture provides a more efficient way of delivering resources for a diverse set of workloads and a platform to do early harwdare-software design explorations. This architecture disaggregates data center resources (primarily CPUs, accelerators, storage, and network interface cards) over a sufficiently fast interconnect fabric. Figure~\ref{fig:comp_concept} shows the high level concept of a composable server architecture. Peripheral Component Interconnect Express (PCI-e)-compatible devices can be plugged in and out in a dynamic fashion.  Such a setup allows for increased flexibility, agility, and efficiency in the usage and maintenance of the data center. The ability to quickly evaluate and determine an optimal system configuration can lead to time to market advantage for any enterprise which adopts this methodology for design and implementation of products. It also serves to decouple the life cycles of the individual resources, allowing for swifter, less expensive adoption of newer technologies at the component level; and therefore, allows for swifter upgrades and refreshes of products. As no specific component is tied to any other, the cost of upgrading individual parts is isolated, and administrators do not have to upgrade unnecessary components. Lastly, the flexible nature allows the data center service a wider range of workloads (e.g., Artificial Intelligence (AI) and Big Data Analytics) while employing a relatively standard set of hardware, and re-configuring it quickly to each workload's requirements. Our primary interest, and the subject of this paper, is the use of a composable server for experimentation of next generation system design with a focus on AI workloads.

Our team, in conjunction with a partner vendor, designed and built a composable infrastructure test bed. In this paper we demonstrate how such as system can be configured with various system-level configurations to evaluate the performance of representative deep learning (DL) benchmarks across vision and Natural Language Processing (NLP) domains and measure the overhead of the system. Our results show that the composable architecture provides great flexibility in dynamically composing a system with desegregated components while introducing acceptable overhead for small and moderately sized DL models. To the best of our knowledge, this is the first work that performs a detailed performance characterization of DL workloads in a composable infrastructure. 

%. As a part of this project, we study the efficacy of the system in letting users carefully tailor the setup of their hosts to efficiently fit their workload needs and product development.

\section{Composable System Overview}
% Lorraine
The composable system consists of the following key elements:%, which will be described in detail in this section, along with how these elements are architected to become a single composable system:
\begin{itemize}
\item Falcon 4016 Composable Chassis
\item Host Servers
\item Management Server
\end{itemize}
\subsection{Concept}

The Falcon 4016 is a 4U GPU chassis which can accommodate up to 16 devices.  These devices can be Graphics Processing Units (GPUs), Network Interface Cards (NICs), Non-Volatile Memory Express (NVMe) Solid State devices (SSD), or custom-developed hardware that is designed to work with a PCI-e 4.0 interface.  The devices in the Falcon can be shared among multiple PCI-e root complexes (servers).  The Falcon has four host ports (H1-H4) which connect the Falcon, via a host adapter, to the host servers, with 400 Gb/s CDFP cables.  The host adapters are a standard, low profile PCI-e 4.0x16 adapter card.  The high speed of these connections between the Falcon and the host servers reduces latency to a minimum between the host server and the Falcon.  

%The Falcon 4016 has a set of LED indicators which indicate standard device status, such as host link status, power status, fan status, system fault, etc.

A graphic user interface (GUI) is provided for administrators and users as a configuration and monitoring interface.  This interface manages the Falcon 4016, including configuring and monitoring the PCI-e switches, devices inside the Falcon, and monitoring of the host servers.

The initial host servers are SuperServer Supermicro SYS-4029GP-TVRT with the following specifications
\begin{itemize}
\item 2 CPUs @ Intel(R) Xeon(R) Gold 6148 CPU @ 2.40GHz 20 Cores
\item 756 GB Memory
\item 8 GPUs @ NVIDIA Tesla V100 SXM2 GPU 16 GB
\item 2 NIC @ Intel Corporation Ethernet Controller 10-Gigabit X540-AT2
\item 1 locally attached NVMe Intel SSDPEDKX040T7 4 TB
\end{itemize}

\subsection{Management Interface}
The management system of the Falcon 4016 is based on the OpenBMC software stack. The OpenBMC project is a widely used Linux distribution for embedded devices. The baseboard management controller (BMC) manages and monitors most of the standard buses in the system, as well as temperature, fan sensors, storage devices, and network. The BMC can alert administrators to any parameters which fall outside of specifications.

The Falcon 4016 can be accessed via a web interface to configure devices, enable operations, and monitor information and provide functional information such as:
\begin{itemize}
\item System Information: model name, serial numbers, last login, etc.
\item Temperature Information: drawers and chassis
\item Resource List: Number of PCI-e links, Host links, GPUs, NVMe SSD devices, etc.
\item GPU Utilization Rate
\item Throughput by drawers and GPU slots
\item PCI-e Link Health, including accumulated error count
\end{itemize}
Additionally, resource information and operations control can be done from the Falcon Management Interface.  There are two views available for this activity - a list view or a topology view.   Views will also depend on the primary role assigned (administrator, user).  A summary of these operations are as follows:
\begin{itemize}
\item View and change topology of the resources, host servers, and PCI-e switch chips%: information that provides details on the devices and their status, and an allocate button for allocating devices to hosts
\item Check resource information, such as device model, link speed, vendor ID, device ID
\item Check resource distribution configuration and port configuration.  The owner of the resource can change port configuration, such as port type and lanes of specific ports
\item Attach/detach resource
\item Check resource status and link capabilities
\item Import or export resource allocation as a configuration file
\item Monitor port traffic, port errors, link speed
\item Define event logs for export (administrator feature)
\end{itemize}
%The administrator can do account management from the Falcon management interface.  All common account actions can be accomplished from this interface (create, delete accounts, edit user groups, monitor users, establish password rules, change passwords, etc.).

\subsection{Composability Features}
In addition to allowing devices to be attached to the hosts, the Falcon 4016 also operates in both standard and advanced mode.  In standard mode, one host can access up to eight GPUs, or two hosts can access up to four GPUs each in a single drawer. In advanced mode, three hosts can share all GPUs.

\subsection{Enterprise Features}
In addition to being an environment which supports R\&D experiments in composability, our composable infrastructure has the added requirement that it be 'enterprise ready'.  There are three key attributes to an enterprise solution: security, scalability and flexibility.  Each of these attributes can be viewed as mechanisms to reduce risk and protect assets, and together create a resilient environment able to operate under stress, and not fail.  'Enterprise ready' is a term that is not often heard in R\&D circles, but is a paramount concern to IT administrators who must run their data centers in strict accordance with large corporate entities' standards and policies for security, compliance, privacy, availability and reliability - all of which are important to customers who invest large amounts of resources on developing intellectual property for competitive advantage, and must protect their investments to stay ahead of competitors. In this day and age, when cybersecurity attacks and thefts have become the norm, no company is willing to leave to chance its most valuable and precious resources. Since our composable environment will be available to our partners for self-service experimentation, via the Internet, this means we must have a robust and safe environment to protect our partners assets.  Our composable server must be able to withstand attack, from external as well as internal sources.  With this in mind, we have developed our environment around a set of enterprise-grade features which will allow users to feel confident that their data and applications are secure.  
\begin{figure}[ht!]
    \centering
    \includegraphics[width=8cm] {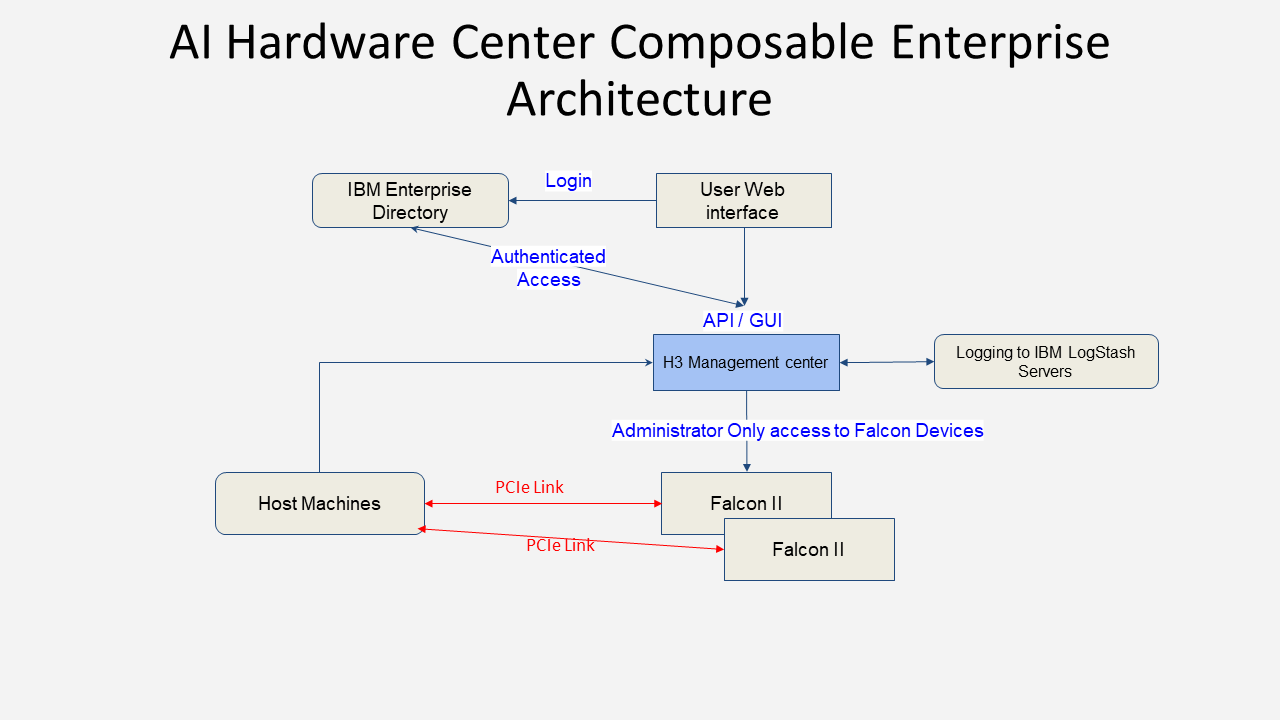}
    \caption{Enterprise Architecture}
    \label{fig:CS_arch}
\end{figure}
In the architecture section above, we described the management environment of the Falcon 4016, which is available to configure the environment.  In production environments, the best practice is not to allow users of the environment to directly access the low level, physical devices.  Access to system services by users can often result in disruption to the entire enterprise, if users are allowed to perform system-level tasks - such as re-assigning devices, disabling / enabling devices, etc.  In order to prevent this from happening in our environment, we worked with our design partner to design and implement a Management Center Server (MCS) which allows users to control their own environment, yet not have any access to other users resources.  By abstracting these services to this higher level, we have an added layer of security which allows users to do productive work on their own, yet not interfere with the work of other users.

\section{Composable System Architecture} 
\label{sec:com-architecture}
\subsection{Topology}
\begin{figure}[ht!]
    \centering
    \includegraphics[width=8cm]{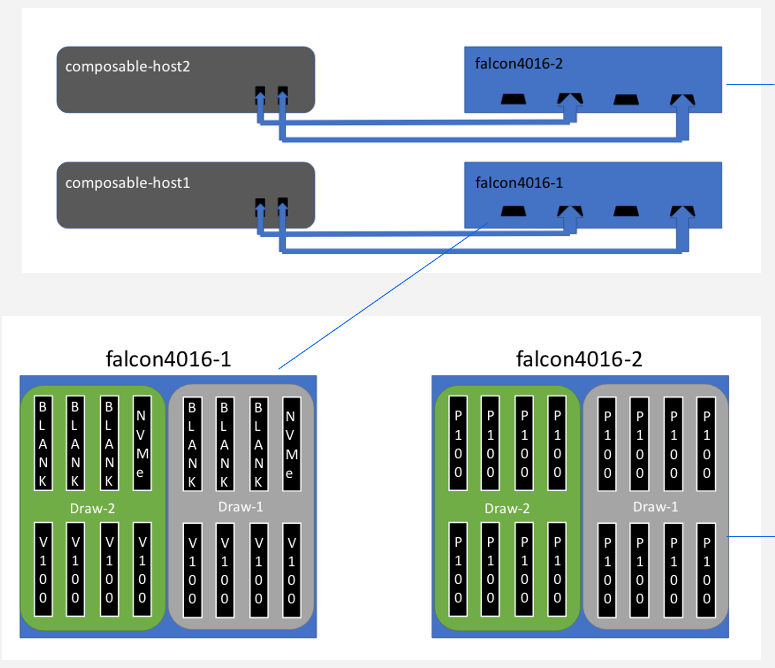}
    \caption{Example High-level Composable Architecture}
    \label{fig:comp_arch}
\end{figure}

The composable systems are made up of the host servers (nodes) and the Falcon 4016 appliances (devices), which extend the resources of the host servers. Figure~\ref{fig:comp_arch} illustrates a sample architecture comprised of two falcon switches. Each of them is connected to a host. By allowing the elements of the system to be varied, the administrator can re-arrange the communication topology of the composable environment.  Modifying the topology allows users to evaluate the impact of the physical topology (i.e., device location, physical interconnections, transmission rates), and the logical topology (data flow within the network).  

\subsection{Modes of Operation}
\begin{figure}[ht!]
    \centering
    \includegraphics[width=8cm]{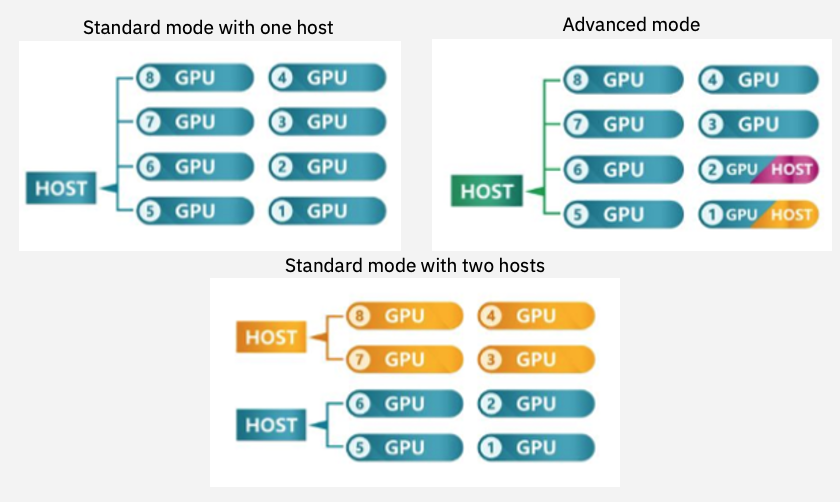}
    \caption{Composable System Modes of Operations}
    \label{fig:falcon-modes}
\end{figure}

Falcon 4016 provides various configuration capabilities to meet the requirements of different usage scenarios. The Falcon 4016 consists of 2 drawers, each with 8 slots, which can be configured independently. There a three modes of operation  as illustrated in Figure~\ref{fig:falcon-modes} with some examples:

\subsubsection{Standard Mode with One Host}
\begin{itemize}
\item One host is connected to the drawer, accessing all eight devices. The two drawers can also both be connected to the same host which gives that host access to all sixteen devices, or can be connected to two different hosts giving each access to eight devices.
\end{itemize}

\subsubsection{Standard Mode with Two Hosts}
\begin{itemize}
\item Two hosts are connected to a drawer with each host accessing four devices.
\item One host can have two connections to the same drawer. Each connection gives access to four devices. This improves performance of communications between host and devices but may slow communications between devices in the two halves of the drawer.
\end{itemize}

\subsubsection{Advanced Mode / Device Dynamic Provisioning}
\begin{itemize}
\item Up to three hosts can be connected to each drawer with the eight devices in the drawer shared among the three hosts in various configurations, depending on how the user configures the setup.
\item Devices can be allocated and re-allocated dynamically on-the-fly across the connected hosts.
\end{itemize}

\subsection{Configurable Features}
The Falcon 4016 has a number of configurable features which allow for running experiments with various devices, numbers of devices, etc. A sampling of the configurable features follows here:

\begin{itemize}
\item Dynamic allocation and de-allocation of devices    
\item Number of hosts connected to the Falcon, as described in the previous section
\item Different types of devices
\end{itemize}

\section{Related Work}
% Kaoutar / Lorraine
The concept of composable systems has been explored in both academic projects, as well as industrial research groups, for many years.  The challenge of deploying the correct mix of CPUs, accelerators, memory, network  and storage so that there is a high system utilization, with few bottlenecks has proven to be an elusive goal. As various authors have pointed out in their research, the key enabler is the network, as well as management software to manage the logical connections of the resources.   This challenge is easily seen in the table below, where the latencies increase from 5x to 100x as the communication paths move from CPU to disk. \cite{7842314} 
\begin{figure}[ht!]
    \centering
    \includegraphics[width=8cm] {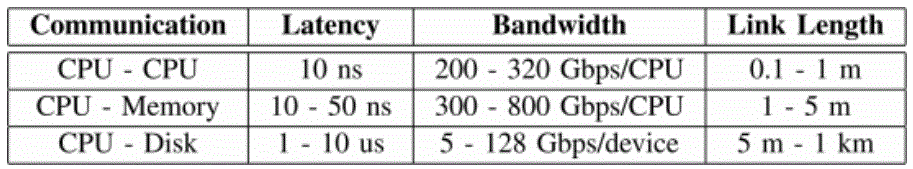}
    \caption{Communications Requirements}
    \label{fig:Comm_reqt}
\end{figure}

In terms of production systems used for industrial and commercial purposes, the concept of dis-aggregation or composability has not made many inroads.  The performance penalties due to latencies introduced by the dis-aggregation of the systems, and the software modifications required to utilize the composable resources, necessitating non-standard hypervisors, or OS changes, have proven to be impediments that are yet to be easily overcome.  Some research experiments have focused on exploiting improvements in Ethernet, which is now moving toward 400 GB/s speeds, with multiple lanes, as well as optical networks.  In \cite{10.1145/2881025.2881030}, the authors focus on the Spark SQL workload to show dis-aggregation is possible, with a commercial network topology such as Ethernet.  The result is reasonable for Spark SQL, but not generally applicable to other workloads, or, possibly, to a very highly scaled implementation of Spark.  
The architecture adopted in traditional data centers based on the interconnection of the Von Neumann architectural elements (compute, memory and network resource) has long been known to have several challenges: low resource utilization, leading to cost and power inefficiencies, slow and difficult upgrades, limited scalability, etc. The rigid architectural boundaries of hardware resources also limits the gains and value of compute virtualization. The composable approach has been explored repeatedly as a means of overcoming these challenges and as a means of accommodating various workloads. A composable infrastructure with high throughput and low latency optical network promises to deliver these requirements. Yet the promise has yet to truly materialize. \cite{Ajibola} Some niche companies have offered products \cite{drivescalewebsite}, as well as larger companies, such as Cisco. \cite{ciscoweb}  However, these offerings have not had significant market impact.

An example of an early exploration is in \cite{Lim1}.  The authors propose the use of a Memory Blade for a commercial blade server, which communicates with client (CPU) blades to handle read-write of memory, memory allocations and address mapping.  The paper proposes two system architecture uses of the memory blade: Page-Swapping Remote Memory, and Fine-Grained Remote Memory Access.  These techniques require changes to the underlying system software, and the improvements do not meet the return on investment that system designers require to adopt specialized system software.  In production, standard operating environments, which can be more easily debugged, and do not require specialized staff knowledge outweigh any small system improvements.  Twelve years later (2021), this still remains true.  In 2012, the same author \cite{Lim2} released a paper that built on the 2009 work.  A  software-based prototype, with Xen hypervisor extensions was developed. The authors demonstrated the usefulness, and some savings for employing memory blades.  The changes to system software, such as replacement policies for pages, memory management code, changes for content-based page sharing have again stymied the use of these techniques in production systems.

Several studies have been done specifically on dis-aggregated memory.  Since memory requires extremely high bandwidth, explorations have centered on the use of optically connected memory in order to employ composability in systems. In  \cite{Abali} the authors explore the use of this technique, in which (Dynamic Random Access Memory) DRAM modules are pooled in chassis and racks.  With fast optical interconnect among compute and memory, the concept is viable at the rack level.  However, latencies create severe performance issues at the data center level, as the cost of the optical interconnect is prohibitive (2015).  A 2016 paper \cite{10.1145/2881025.2881030} explored the use of dis-aggregated memory for Spark SQL analytic queries. The results showed promise for this one workload using available technology.   In 2017 \cite{Meyer} explored memory dis-aggregation and the development of a simulation framework for evaluation.  The technique in this exploration was to profile information of data center applications that fed into the simulator developed for this work.  With a mean error of 10\% percent, simulation results demonstrate that high packetization times (packetization is required for remote memory access) may degrade application performance up to 66 \% and low memory access bandwidths can degrade up to 20\% of the performance.  

A recent paper which explored composable memory is \cite{8732926}.  In these experiments, the authors propose the use of a RAM-based file system (temporary storage, tmpfs), SCSI storage protocol, RDMA over Converged Ethernet (RoCE), and OS (Virtual Machine Monitor) VMM to create a production level composable memory.  These experiments advance the research on composable memory, but bottlenecks inside the OS, and latencies are still problems for widespread adoption.

In 2018 the authors of \cite{ihsin-2019} described and reviewed a composable system experimental platform which they developed for research.  The authors performed experiments with accelerators, storage devices and network devices.  The paper focused on reporting results of performance and reliability.  The authors also introduced the usefulness of the composable system in hardware/software co-design, which we are focusing on in our experimentation and research.  In \cite{LI2017180} we see the authors focusing on a rack scale composable system using PCI-e switches, for use in a rack scale architecture.  The concept has conceptual merit for production systems, but is yet to be fully implemented in commercial systems.  

A 2020 paper introduced the concept of Rack Memory (RackMem) which the authors describe as an efficient implementation of dis-aggregated memory for rack scale computing. \cite{10.1145/3410463.3414643}  This research focuses on improving the Linux demand paging algorithm and demand adjustment to remote memory access.  RackMem uses local memory as a cache, and with the configuration of operations in user space at run-time, is able to implement application specific caching policies which allow the applications to take advantage of the distributed/dis-aggregated memory. RackMem achieves higher throughput for a wide variety of workloads, with changes to the Linux operating environment's demand paging, and RDMA implementation.  

Our work takes the approach described in  \cite{ihsin-2019}.  We have built a composable system, based on standard hosts, and a composable appliance that offers hardware and software co-design, allowing for exploration of varying numbers of accelerators, IO cards, solid state storage, in order to determine the optimal configuration prior to final commitment of system build.  In this way, we are making use of the composable infrastructure for insights into new production environments, helping to reduce 'guesswork' on final build out.

\section{Performance Analysis}
In this section we detail the performance characterization of different DL training benchmarks on the composable system with various system-level configurations. We briefly outline the various configurations that we have used for the performance characterization. We also study how each benchmark is exercising different components of the system and the impact of the PCI-e switching on the various benchmarks. 

\subsection{Experimentation Setup}
\subsubsection{Hardware Environment}
\begin{figure}
    \centering
    \includegraphics[width=8cm]{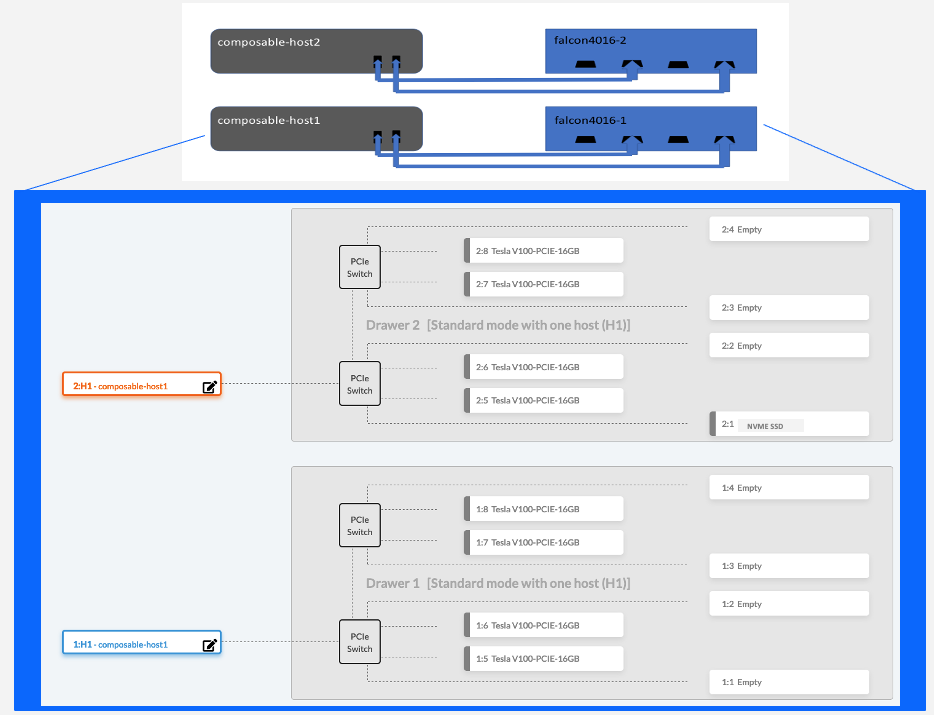}
    \caption{Composable System Topology Used in the Experimental Evaluation}
    \label{fig:setup_top}
\end{figure}

\begin{figure}
    \centering
    \includegraphics[width=8cm]{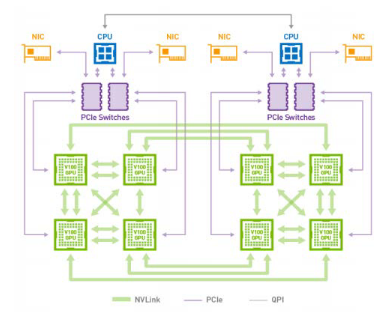}
    \caption{Hybrid Cube Mesh NVLink Topology (Source: NVIDIA website)}
    \label{fig:nvlink-v100}
\end{figure}

As described in Section~\ref{sec:com-architecture}, our composable infrastructure consists of 2 Falcon 4016 systems. Each Falcon system is connected to a Supermicro host, NVIDIA Tesla V100 and P100 GPUs, and NVMe SSD drives. Each host has 8 local Tesla V100 GPUs.  For the purpose of the experimental evaluation, we have used a subset of this system to evaluate the DL benchmarks. Figure~\ref{fig:setup_top} shows the topology of the system used. The host is connected to both drawers in the Falcon. Each drawer contains 4 Tesla V100 GPUs. Drawer 2 also has a 4TB NVMe storage device attached to it. We also have an additional 4TB NVMe storage device that we attached to the host directly. This setup gives the total of 16 GPUs, a Falcon-attached NVMe drive, and a locally attached NVMe drive. The GPUs attached to the Falcon have 16 GB High Bandwidth Memory 2 (HBM2) GPU memory and use the PCI-e switching fabric for GPU-to-GPU communication. The Tesla V100 GPUs that are locally attached to the host also have a 16GB HBM2 GPU memory but have an NVLink SMX2 fabric for GPU-to-GPU communication with a maximum speed of 300GB/s. Second-Generation NVIDIA NVLink™ provides significantly more performance for both GPU-to-GPU and GPU-to-CPU system configurations compared to using PCI-e interconnects. Figure~\ref{fig:nvlink-v100} illustrates the hybrid cube mesh NVLink topology used in the host. The Falcon systems add flexibility to the system, allowing multiple configurations of resources.

\subsubsection{Software Environment}
To ensure reproducibility of our results, Table~\ref{tab:sw-stack} lists the details of the software stack that we have used in our evaluation. PyTorch is a popular deep learning framework. The NVIDIA CUDA library is a parallel  programming interface to NVIDIA GPUs. The NVIDIA’s cuDNN (deep neural network) library provides device-level optimized implementation for neural-network back-end kernels. The NVIDIA NCCL (collective communication) library provides a multi-GPU communication interface that supports several communication fabrics, such as NVLink, PCI-e, and Ethernet. We used Weights \& Biases~\cite{wandb} for DL benchmark experiments, tracking and collecting of various system-level metrics. We also used the NVIDIA Nsights~\cite{nvidia-profilers} profiling tools for GPU and CPU sampling and tracing and GPU kernel profiling.

\begin{table}[h]
    \centering
    \begin{tabular}{|c|c|}
        \hline
         Operating system & Ubuntu 18.04 \\ \hline
         DL Framework   &  PyTorch 1.7.1 \\ \hline 
         CUDA    &  10.2.89 \\ \hline 
         CUDA Driver & 450.102.04 \\ \hline
         CUDNN        &  cudnn7.6.5 \\ \hline 
         NCCL      &  NCCL 2.8.4\\ \hline %2.7.8
         Profilers & wandb 0.10.14  \\ 
                   & NVIDIA Nsight Systems 2020.4.3.7 \\
                   & NVIDIA Nsighy Compute 2020.3.0.0 \\ \hline
    \end{tabular}
    \caption{Software Stack Details}
    \label{tab:sw-stack}
\end{table}

\subsection{Deep Learning Benchmarks}
\begin{figure}[ht!]
    \centering
    \includegraphics[width=8cm]{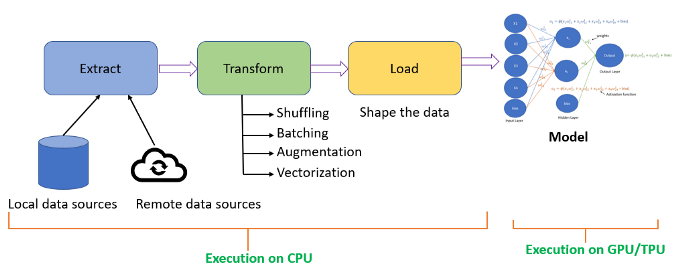}
    \caption{Deep Learning Training Data Workflow}
    \label{fig:data-workflow}
\end{figure}

We have selected several representative DL benchmarks to cover various ranges of model size, computation to communication ratio, pre-processing steps, application modalities, and different types of neural network architectures. Training DL benchmarks follows a data driven approach that uses the stochastic gradient descent approach. The training goes through multiple iterations of feeding data to the model and adjusting its parameters to reduce the training loss. During each iteration, a batch of data is loaded from the hard drive to the host memory, then the data is transferred to GPU memory through the PCI-e bus where most of the DL computation happens. In certain cases, the batch of data undergoes some pre-processing in CPU memory before being transferred to GPU memory. If multiple GPUs are used, the common technique used is the data parallel approach, where each GPU gets an exact copy of the neural network and applies the same computation on different sampled batches. After every iteration, the GPU synchronize using a collective communication mechanism. In all our experiment we are using the NVIDIA Collective Communications Library (NCCL) all-reduce collective communication implementation. Figure~\ref{fig:data-workflow} shows the general data workflow of a typical DL workload. 

%TODO: replace this figure with my own drawing
Table~\ref{tab:DL-benchmarks} shows the list of representative benchmarks chosen for this work and their model characteristics. We have selected the ResNet, MobileNet, and Yolo5 from the computer vision domain and BERT from the NLP domain. These are very popular benchmarks that have had wide adoption. This set also represents a wide range of characteristics in terms of depth, number of parameters and the types of neural networks used. Computer vision models typically stress the memory footprint and the inter-device communication costs, while the NLP models are more stressful to the GPU computation time and GPU memory footprint.

\begin{table}[h]
    \centering
    \begin{tabular}{|c|c|c|c|c|}
        \hline
         Benchmarks & Domain & Dataset & Parameters & Depth \\ \hline
         MobileNetV2 & Computer Vision & ImageNet & 3.4M  & 53 \\ \hline
         ResNet-50   & Computer Vision & ImageNet & 25.6M & 50 \\ \hline 
         YOLOv5-L    & Computer Vision & Coco     & 47M   & 392 \\ \hline 
         BERT        & NLP (Q\&A)      & SQuAD v1.1 & 110M  & 12 \\ \hline 
         BERT-L      & NLP (Q\&A)      & SQuAD v1.1 & 340M  & 24 \\ \hline 
    \end{tabular}
    \caption{Characteristics of The Evaluated Deep Learning Benchmark}
    \label{tab:DL-benchmarks}
\end{table}

\subsubsection{Computer Vision Models}
ResNet~\cite{resnet-2016}, short for Residual Networks, is a popular neural network that is used as the backbone for many computer vision tasks. ResNet won the ImageNet Large Scale Visual Recognition Challenge~\cite{imagenet-2015} breaking the record of its predecessor, AlexNet~\cite{alexnet-2012}, and further reducing the top-five error rates from 16.4\% to 3.57\%. One of the fundamental breakthroughs that this model invented is the residual connection which allowed training extremely deep neural networks with 150+layers successfully and solved the classical vanishing gradient problem. 

MobileNetV1~\cite{mobilenetv1-2017} represents a family of computer vision models that are designed for mobile devices. They exhibit a fewer number of operations to cope with the low latency and stringent storage requirements of mobile devices . MobileNetv2~\cite{mobilenetv2-2018} builds upon the ideas of MobileNetV1 through its use of depth-wise separable convolution as efficient building blocks. V2 introduced two new features to the architecture: linear bottlenecks between the layers and shortcut connections between the bottlenecks. These enhancements made the V2 model faster with 2x fewer operations and 30\% fewer parameters. 

YOLO, short for \textit{You Only Look Once}, is another popular vision network that has revolutionized real-time object detection since its first version, introduced in 2016~\cite{Redmon2016YouOL}. YOLO uses a convolutional neural network (CNN) for doing object detection in real-time. The algorithm applies a single neural network to the full image, and then divides the image into regions and predicts bounding boxes and probabilities for each region. These bounding boxes are weighted by the predicted probabilities. One of the biggest advantages of YOLO is its speed with at least 45 frames/s which is better than real time. This network was inspired by the GoogleNet~\cite{googlenet-2014} model for image classiﬁcation. Instead of the inception modules used by GoogleNet, YOLO simply uses 1×1 reduction layers followed by 3×3 convolutional layers. It has 24 convolutional layers followed by 2 fully connected layers. We are using YOLOv5, a recent PyTorch implementation by Ultralytics with higher interference speed than most of the prior detectors.

\subsubsection{Natural Language Processing Models}
Natural language processing deals with building computational algorithms to automatically analyze and represent human language. Similar to computer vision, NLP is one of the most successful application domains for DL across a variety of NLP tasks,  such as machine translation and question answering. BERT (Bidirectional Encoder Representations from Transformers) is a benchmark in a recent paper published by researchers at Google~\cite{devlin-etal-2019-bert}. BERT is considered a breakthrough in the NLP landscape as it presented state-of-the-art results in a wide variety of NLP tasks such as Question Answering (SQuAD v1.1), Natural Language Inference (MNLI), and others. BERT is based on the transformer architecture and attention mechanism~\cite{NIPS2015_7137debd}. It is bidirectional which implies that it learns information from both the left and right sides of the token's context during the training phase. BERT follows a 2-step process:
\begin{itemize}
\item Pre-training phase: BERT learns from a large corpus of unlabeled data including Wikipedia (2,500 million words) and Book Corpus (800 million words). It uses the semi-supervised sequence learning approach by masking out a random word in a sentence. The resulting model is called a \textit{base-model}
\item Fine-tuning phase: In this phase, the \textit{base-model} is fine-tuned to a specific NLP task. This is accomplished by adding a classification Linear layer designed for the specific task and training the model using a labeled and much smaller data-set compared to the large corpora that was used for the initial pre-training phase. 
\end{itemize}
We use the pre-trained BERT-base and BERT-large models and fine tune them on the SQuAD dataset. SQuAD, short for the Stanford Question Answering Dataset~\cite{squad-2016}, is a Question Answering task that consists of more that 100,000 pairs of questions and the corresponding relevant paragraph that contains the answer. 

\subsection{Performance Evaluation}
In this section, we detail the performance analysis of the described DL benchmarks using 5 different GPU and storage configurations of the composable system as described in Table~\ref{tab:exp-configs}. We then measure various system-level metrics such as CPU, GPU, Memory, and PCI-e traffic. 
\begin{table}[h]
    \centering
    \begin{tabular}{|c|c|}
        \hline
         Label & Host Configuration \\ \hline
         localGPUs  & 8 local GPUs and local storage \\ \hline 
         hybridGPUs & 4 local GPUs, 4 falcon GPUs, and local storage\\ \hline 
         falconGPUs & 8 falcon-attached GPUs \\ \hline
         localNVMe  & 8 local GPUs and local NVMe \\ \hline 
         falconNVMe & 8 local GPUs and falcon-attached NVMe\\ \hline 
    \end{tabular}
    \caption{Composable Host Configurations}
    \label{tab:exp-configs}
\end{table}

\begin{figure}[t]
     \centering
     \begin{subfigure}[b]{0.24\textwidth}
         \centering
         \includegraphics[width=\textwidth]{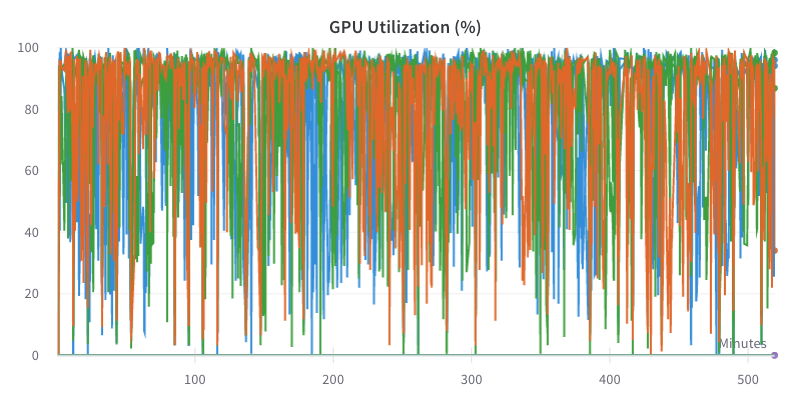}
         \caption{MobileNetV2}
         \label{fig:MobileNetV2-long}
     \end{subfigure}
     \begin{subfigure}[b]{0.24\textwidth}
         \centering
         \includegraphics[width=\textwidth]{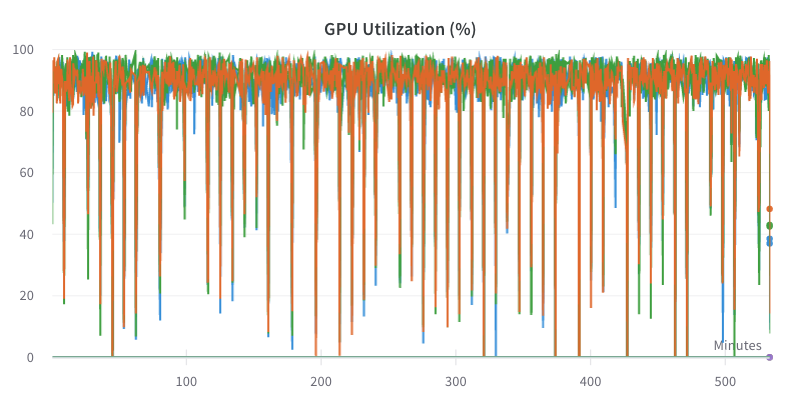}
         \caption{ResNet-50}
         \label{fig:ResNet-50-long}
     \end{subfigure}
     \begin{subfigure}[b]{0.24\textwidth}
         \centering
         \includegraphics[width=\textwidth]{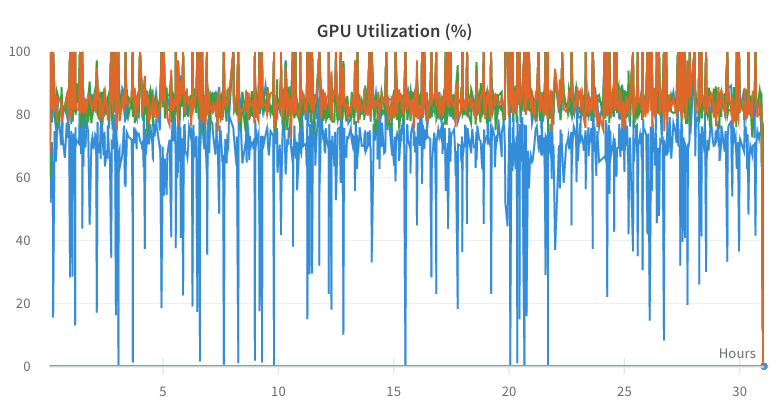}
         \caption{YOLOV5}
         \label{fig:Yolo-long}
     \end{subfigure}
     \begin{subfigure}[b]{0.24\textwidth}
         \centering
         \includegraphics[width=\textwidth]{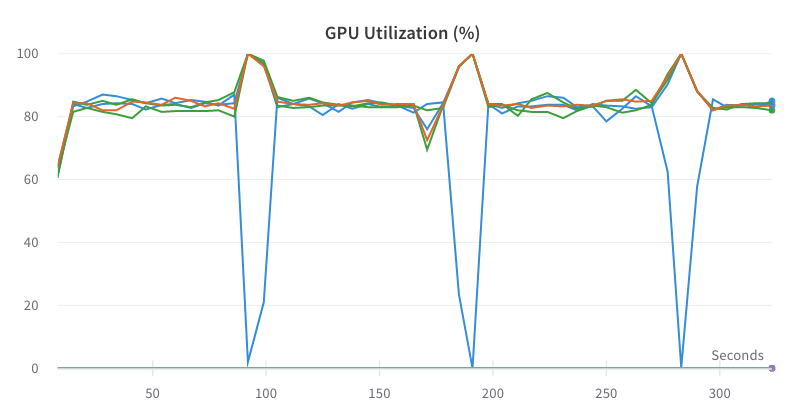}
         \caption{BERT-base}
         \label{fig:bert-long}
     \end{subfigure}
     \begin{subfigure}[b]{0.24\textwidth}
         \centering
         \includegraphics[width=\textwidth]{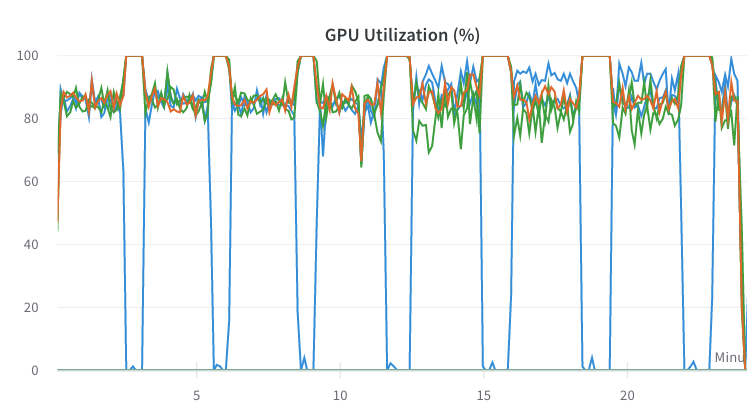}
         \caption{BERT-large}
         \label{fig:bertl-long}
     \end{subfigure}
        \caption{GPU Utilization Patterns for the DL Benchmarks}
        \label{fig:gpu-timeline}
\end{figure}

\subsubsection{GPU and CPU Performance}

\begin{figure}[t]
     \centering
     \begin{subfigure}[b]{0.24\textwidth}
         \centering
         \includegraphics[width=\textwidth]{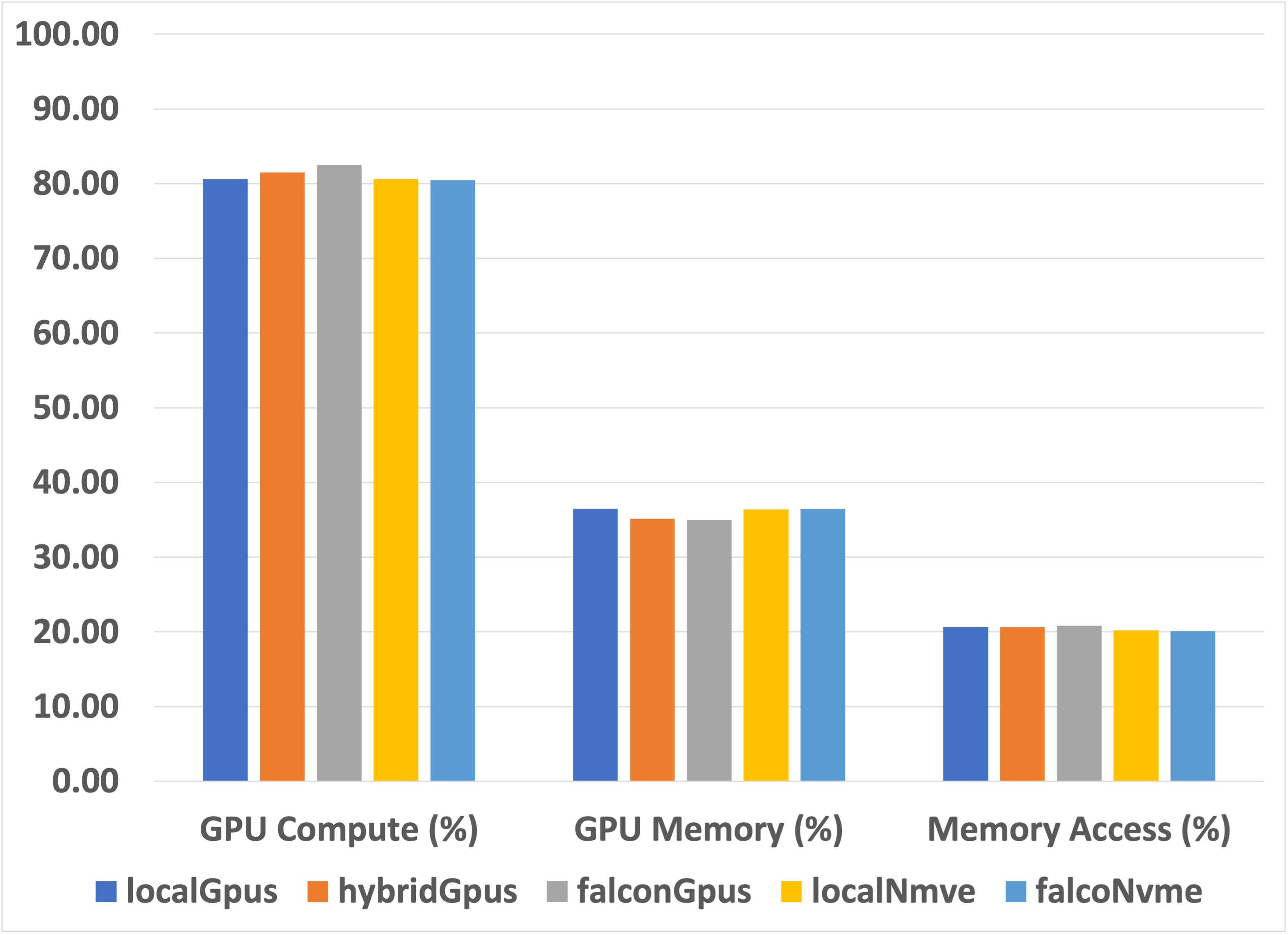}
         \caption{MobileNetV2}
         \label{fig:MobileNetV2-gpu}
     \end{subfigure}
    % \hfill
     \begin{subfigure}[b]{0.24\textwidth}
         \centering
         \includegraphics[width=\textwidth]{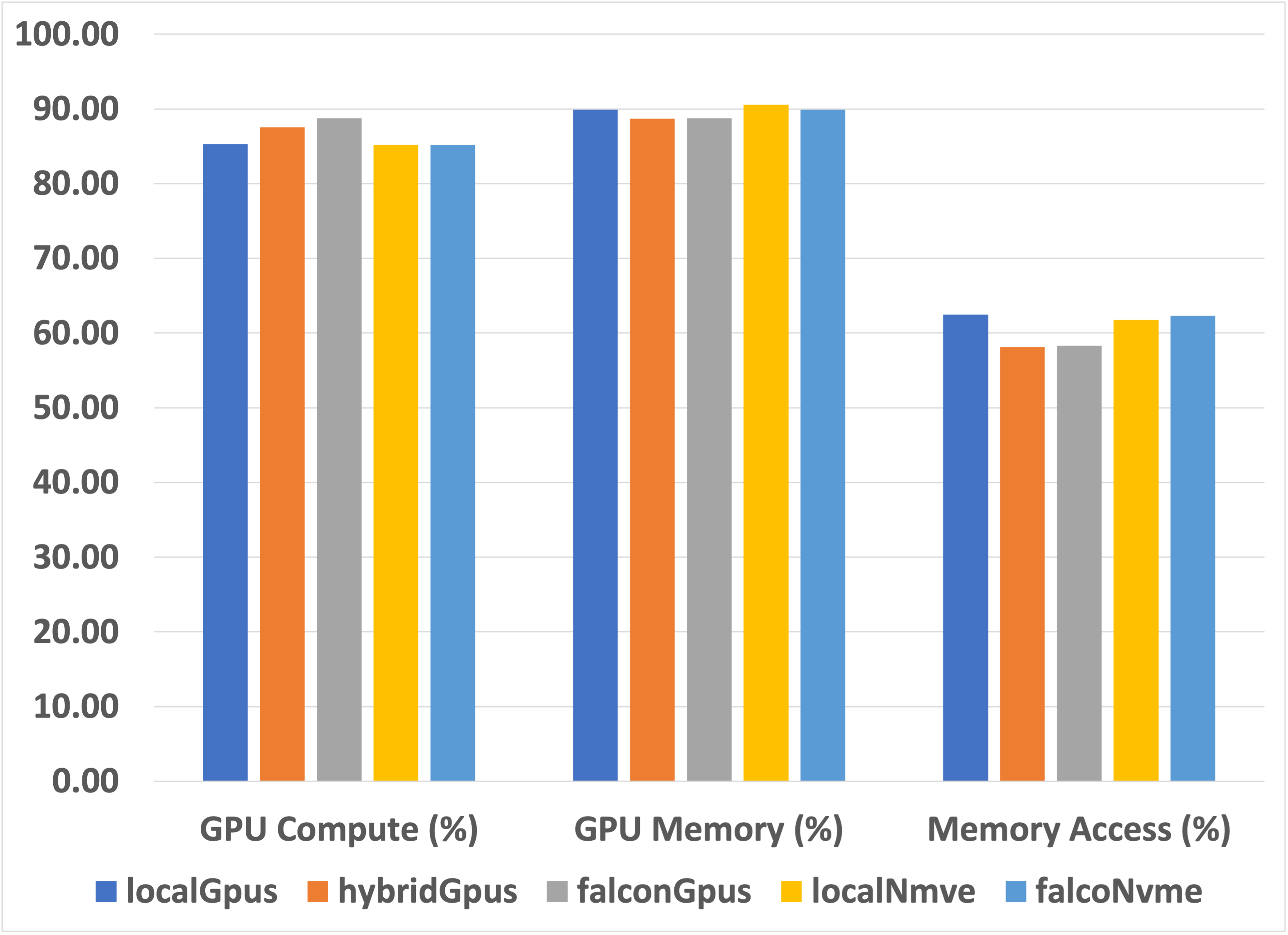}
         \caption{ResNet-50}
         \label{fig:ResNet-50-cpu}
     \end{subfigure}
     %\hfill
     \begin{subfigure}[b]{0.24\textwidth}
         \centering
         \includegraphics[width=\textwidth]{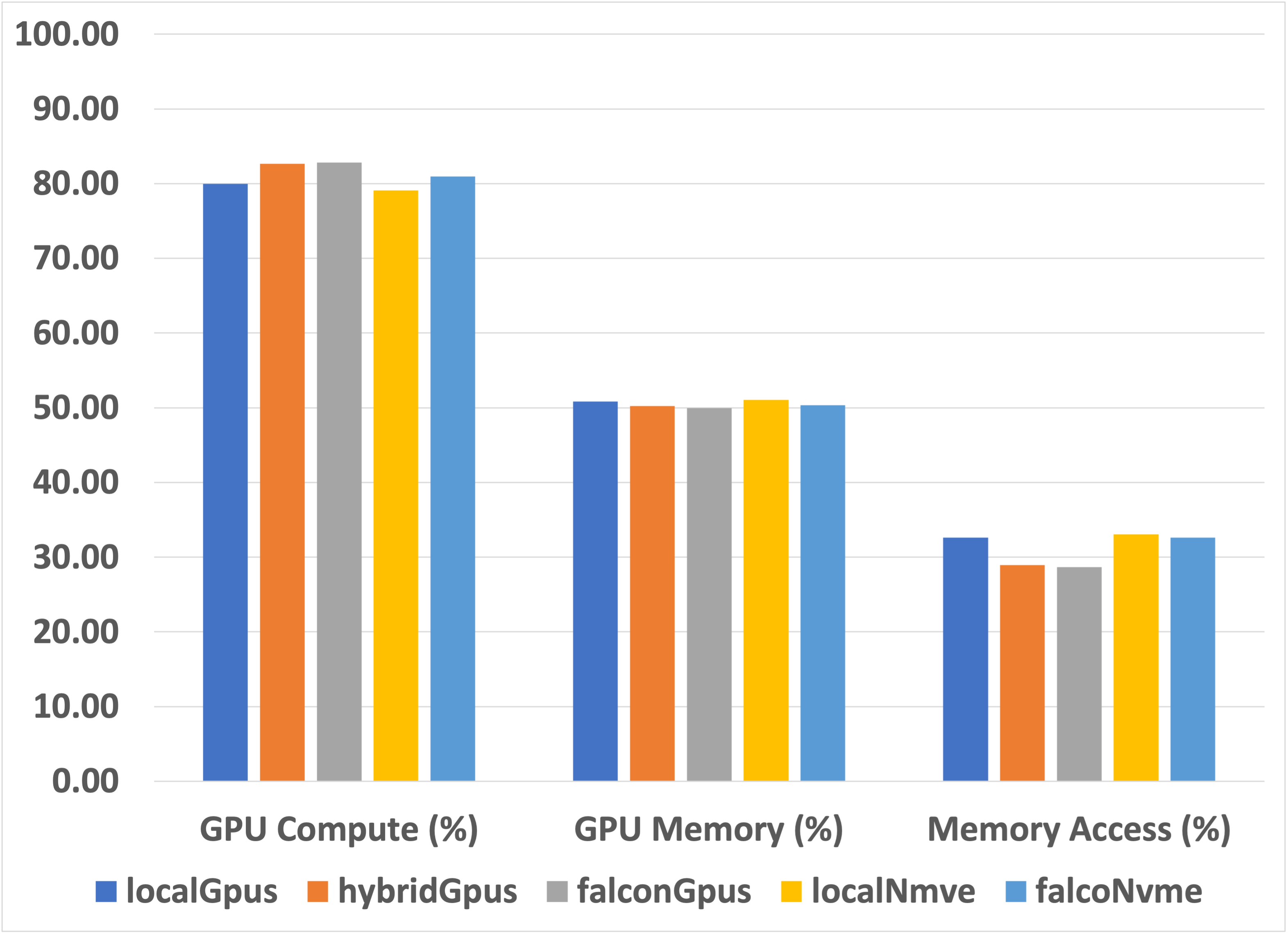}
         \caption{YOLOV5}
         \label{fig:Yolo-gpu}
     \end{subfigure}
     \begin{subfigure}[b]{0.24\textwidth}
         \centering
         \includegraphics[width=\textwidth]{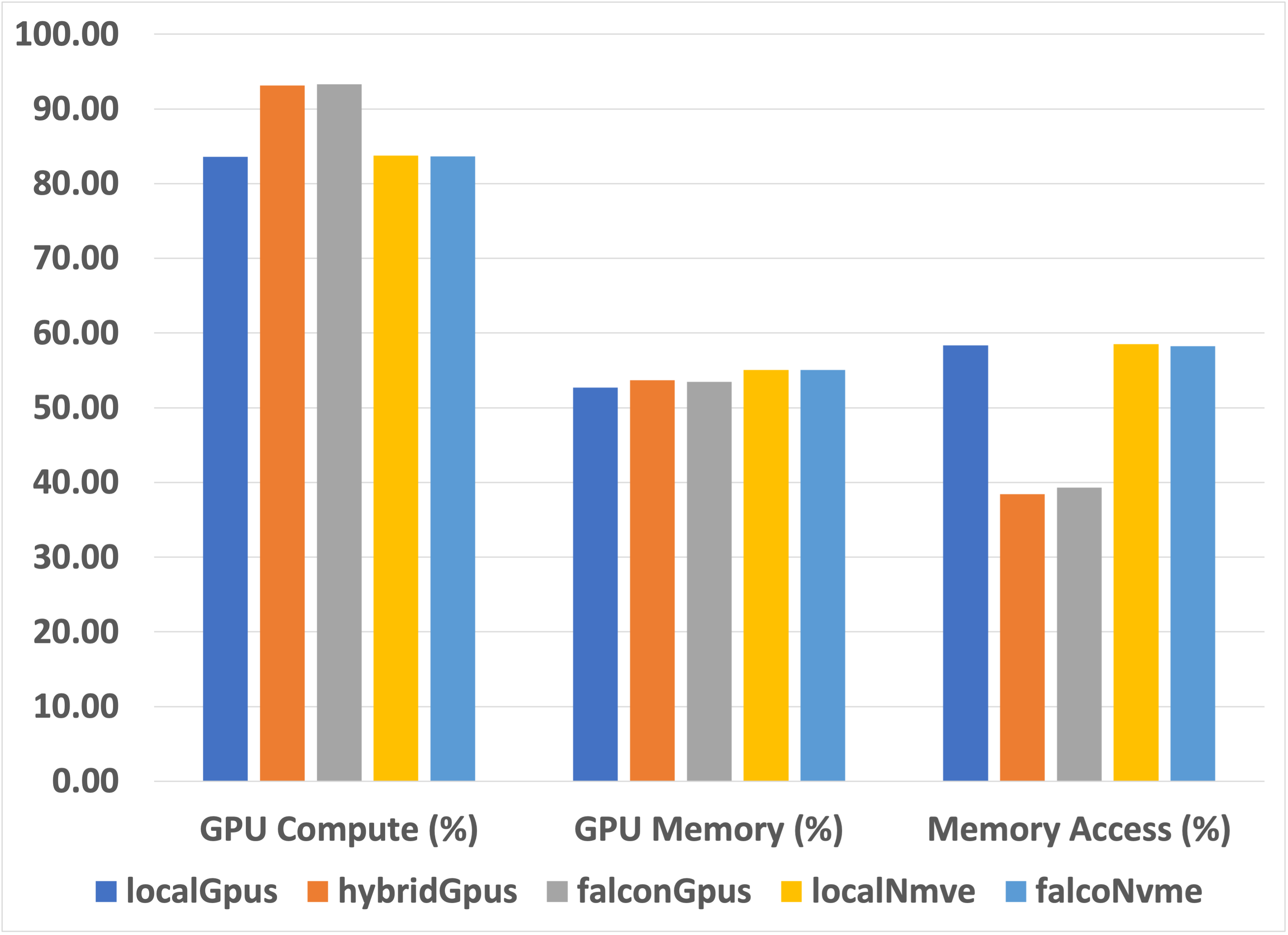}
         \caption{BERT-base}
         \label{fig:bert-gpu}
     \end{subfigure}
     \begin{subfigure}[b]{0.24\textwidth}
         \centering
         \includegraphics[width=\textwidth]{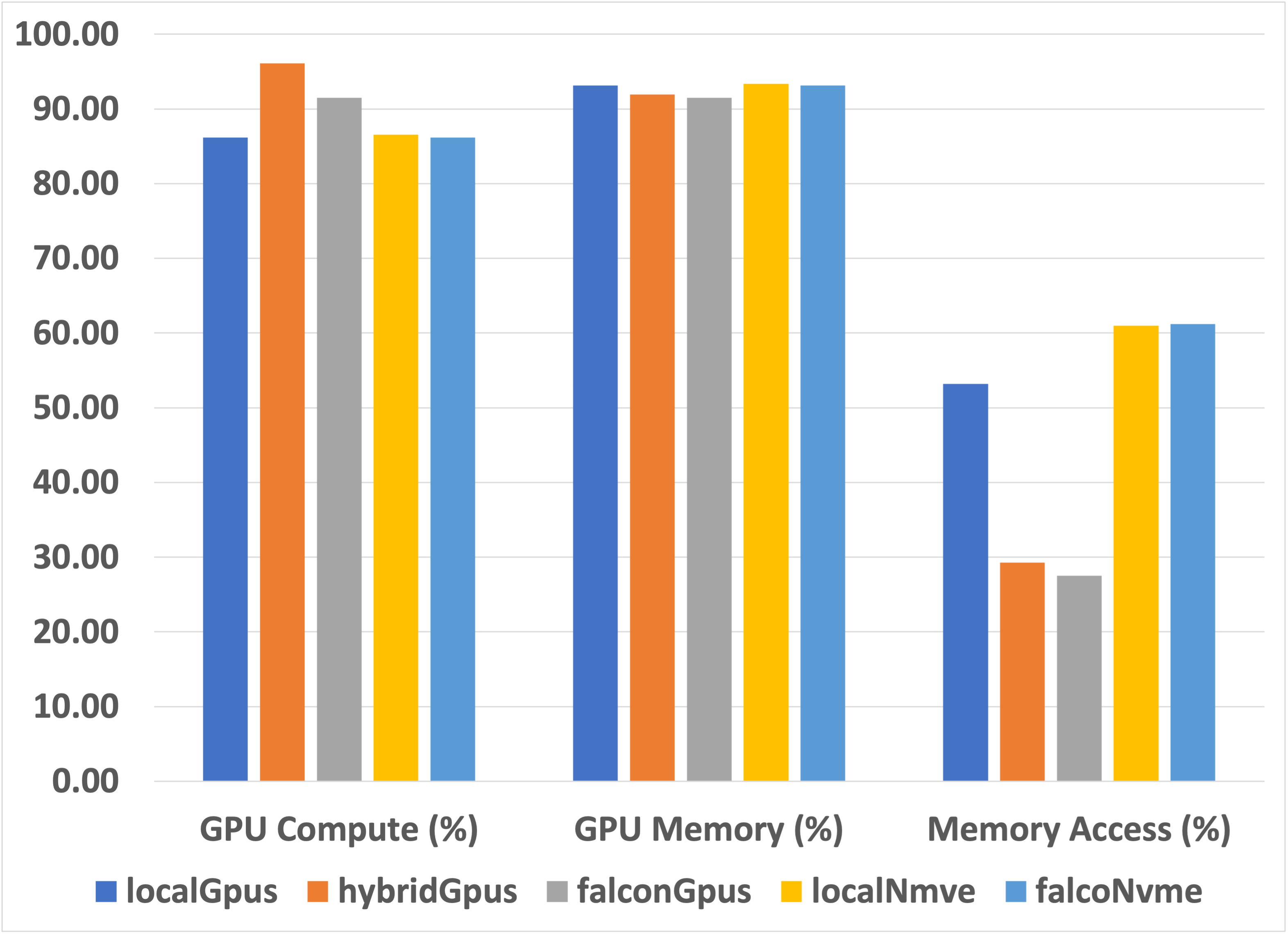}
         \caption{BERT-large}
         \label{fig:bertl-gpu}
     \end{subfigure}
        \caption{GPU Performance of the DL Benchmarks on the Composable Configurations}
        \label{fig:gpu-perf}
\end{figure}
We measure the computation performance of the studied DL benchmarks in terms of the speed of training the models. We have trained all benchmarks up to conversion on the local GPUs configuration to get a sense about the GPU performance profile for each benchmark. As shown in Figure~\ref{fig:gpu-timeline}, all the models have a repeating pattern of exercising the GPUs. All the models use PyTorch optimized Distributed Data Parallel implementation(DDP) with NVIDA NCCL backend. In this approach, multiprocessing is used to duplicate the model across multiple GPUs, each of which is controlled by one process. Every process does identical tasks, and each process communicates with all the others. Only gradients are passed among the processes/GPUs to minimize the amount of data communicated. 
%During training, each process loads its own minibatches from disk and passes them to its GPU. Each GPU does its own forward pass, and then the gradients are all-reduced across the GPUs. Gradients for each layer do not depend on previous layers, so the gradient all-reduce is calculated concurrently with the backwards pass to further alleviate the networking bottleneck. At the end of the backwards pass, every node has the averaged gradients, ensuring that the model weights stay synchronized.
Since some of these benchmarks take a considerable time to train and because of the repetitive nature of the model training, we have elected to train the models for fewer number of epochs to evaluate how the training is exercising the various components of the composable system across the multiple system configurations. We run YOLOv5 for 20 epochs and a batch size of 88, ResNet-50 for 20 epochs and 128 batch size, and MobileNetv2 for 10 epochs and 64 batch size. For BERT-base, we used a max sequence length of 384, a batch size of 96 and 2 epochs. For BERT-large, we fine tuned BERT-large for 2 epochs with 48 batch size and 384 sequence length.
Figure~\ref{fig:gpu-timeline} shows the progression of the GPU utilization across the complete training runs. We see that some benchmarks, like BERT-base and Bert-large, are using the GPU more effectively than other benchmarks. There are sharp periodic drops of some of the GPU's utilization that is mostly attributed to periodic synchronization and check pointing of the models. 

\subsubsection{Impact of Falcon Switching}
\begin{figure}
    \centering
    \includegraphics[width=8cm]{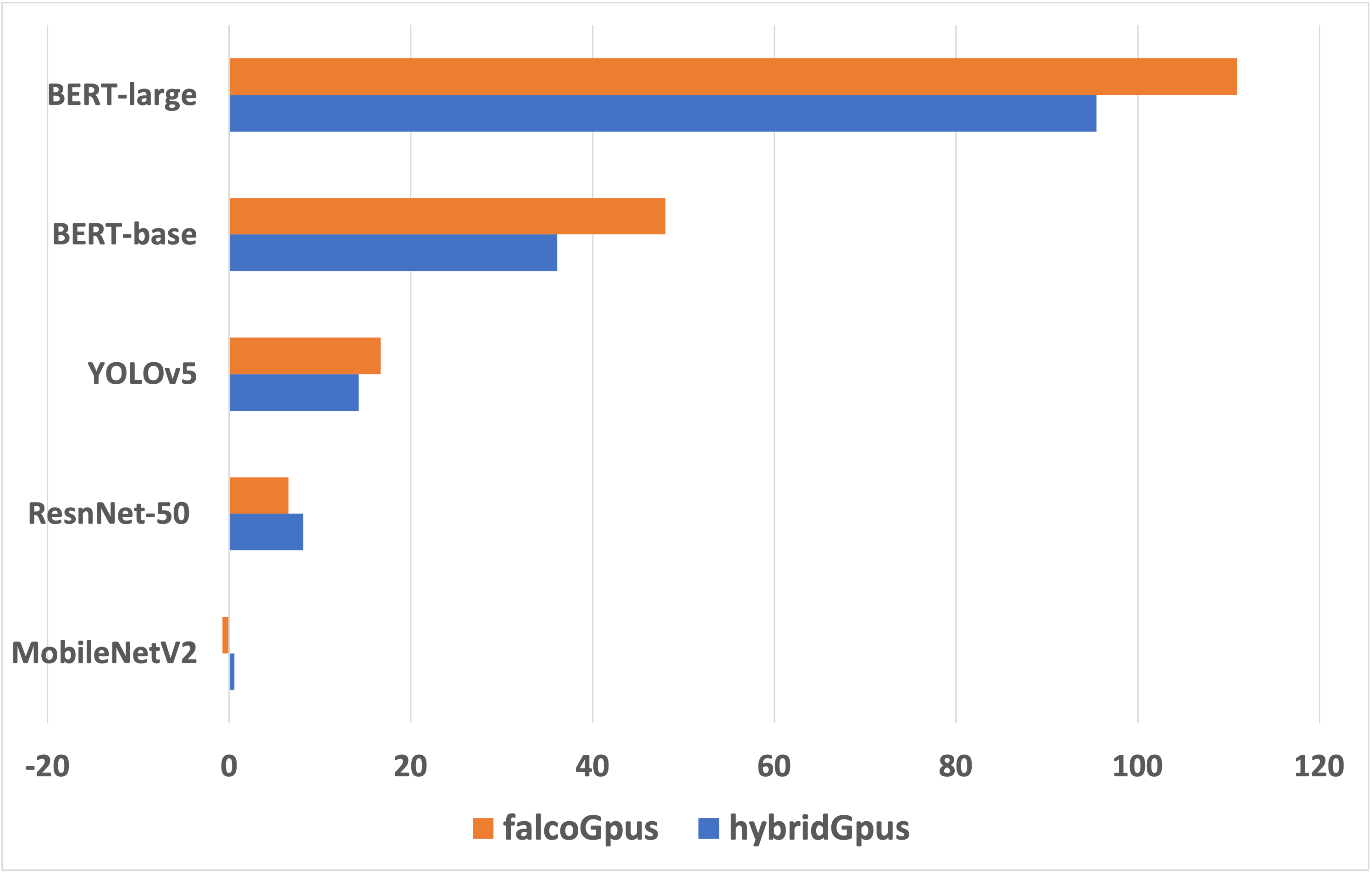}
    \caption{Percentage Change of Training Time from the Local GPUs Configuration}
    \label{fig:runtime}
\end{figure}

To evaluate the impact of the PCI-e switching overhead introduced by the Falcon, we trained all the models on the localGPUs, falconGPUs and hybridGPUs system configurations as described in Table~\ref{tab:exp-configs}. The GPUs that are attached to the host through the falcon switch, use PCI-e Gen4 links, and also communicate among themselves through PCI-e express links, while the local GPUs use the NVLink protocol. Table~\ref{tab:nccl-tests} shows the GPU-to-GPU latency and bandwidth in the various configurations: Falcon to Falcon (F-F), Local to Falcon (L-L) and Falcon to Local (F-L). We notice that the L-L bandwidth is almost 4x faster than the F-L and almost 3x faster than the F-F.

\begin{table}[h]
    \centering
    \begin{tabular}{|c|c|c|c|}
        \hline
          & L-L & F-L & F-F \\ \hline
         Bidirectional Bandwidth (G/s) & 72.37 & 19.64 & 24.47\\ \hline 
         P2P Write Latency (us) & 1.85 & 2.66 & 2.08\\ \hline 
         Link Protocol & NVLink & PCI-e 4.0 & PCI-e 4.0 \\\hline
    \end{tabular}
    \caption{GPU-GPU Bandwidth, Latency, and Protocol}
    \label{tab:nccl-tests}
\end{table}

\begin{figure}[ht!]
    \centering
    \includegraphics[width=8cm]{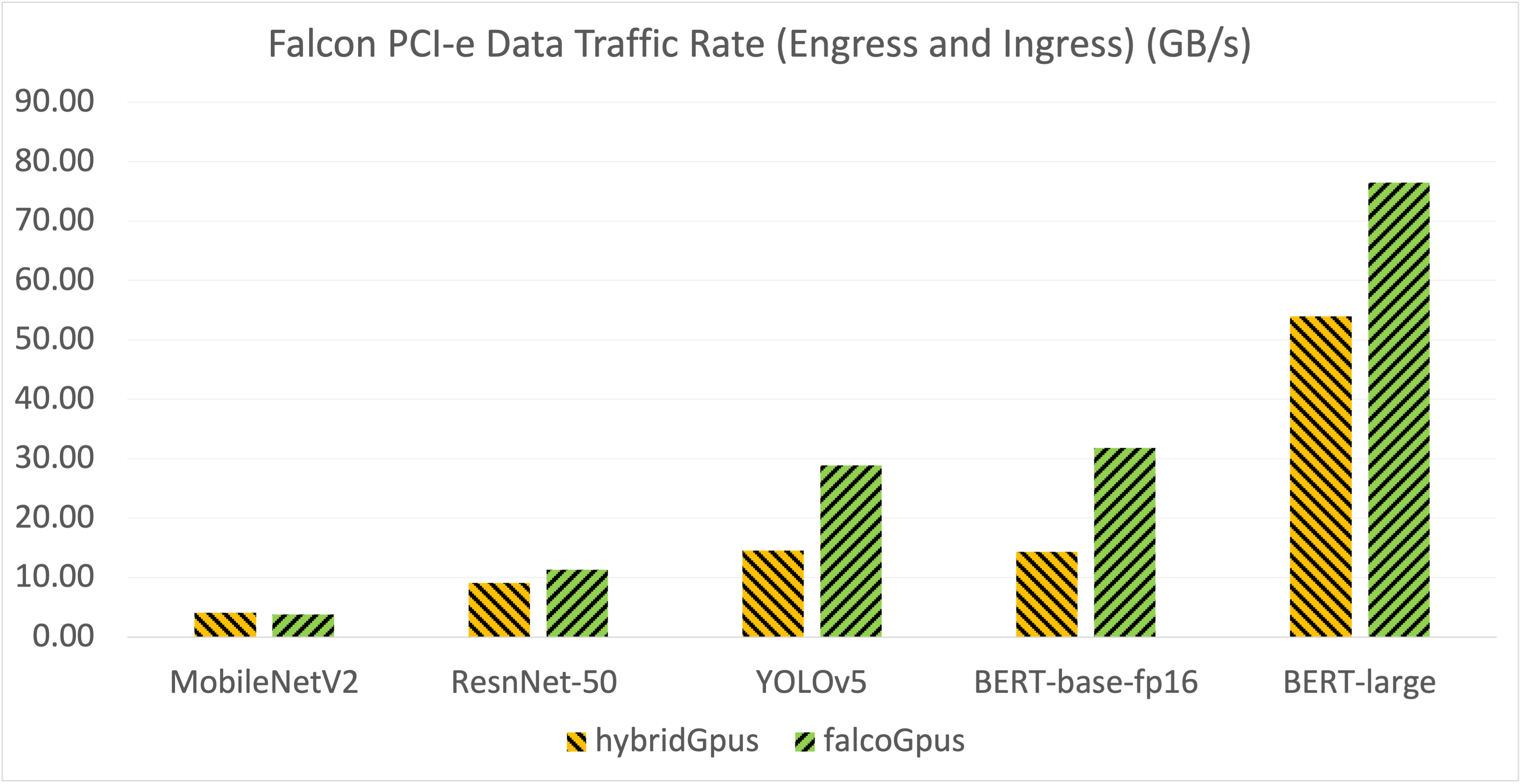}
    \caption{PCIe Data Transfer Rate (GB/s) for Falcon-attached GPU Configurations}
    \label{fig:pcie-traffic}
\end{figure}

Figure~\ref{fig:runtime} shows the relative training time compared to using the host with local GPUs for all DL benchmarks. As the mode size or number of model parameters increases, we start noticing the impact of the PCI-switching. In such cases, the models need to exchange very large amounts of data and the inter-GPU communication can become a bottleneck. For smaller models, such as MobileNetv2 and ResNet-50, the overhead of the PCI-e switching is negligible. The speed of training is less than 5\% slower than the local GPUs configuration. Overall for the vision workloads, the training is less than 7\% slower when using a GPU configuration that involves the Falcon. However, for the NLP workloads we notice that there is overhead introduced by the PCI-e switching and this mostly contributed to the large number of parameters these models have. We can see the correlation between the overhead and the size of the model. BERT-large fine-tuning time took almost twice as much time using Falcon-attached GPUs. This model has 340 million parameters (13x of ResNet-50's parameters). Figure~\ref{fig:pcie-traffic} shows the average amount of data that has been exchanged with the Falcon-attached GPUs using hybridGPUs and falconGPUs configurations. We measured this data by calculating the data transferred every second through the Ingress ports and Egress ports for the PCIe links attached to the Falcon GPUs. As the models become larger, the data exchanged between the GPUs is significantly larger. For instance, PCIe traffic for BERT-large fine-tuning on the falcon GPUs configuration was $76.43 GB/s$, about 19x the PCIe traffic of MobileNetv2 ($4 GB/s$) and 7x of ResNet-50 ($11.31 GB/s$).

\begin{figure}[ht!]
    \centering
    \includegraphics[width=8cm]{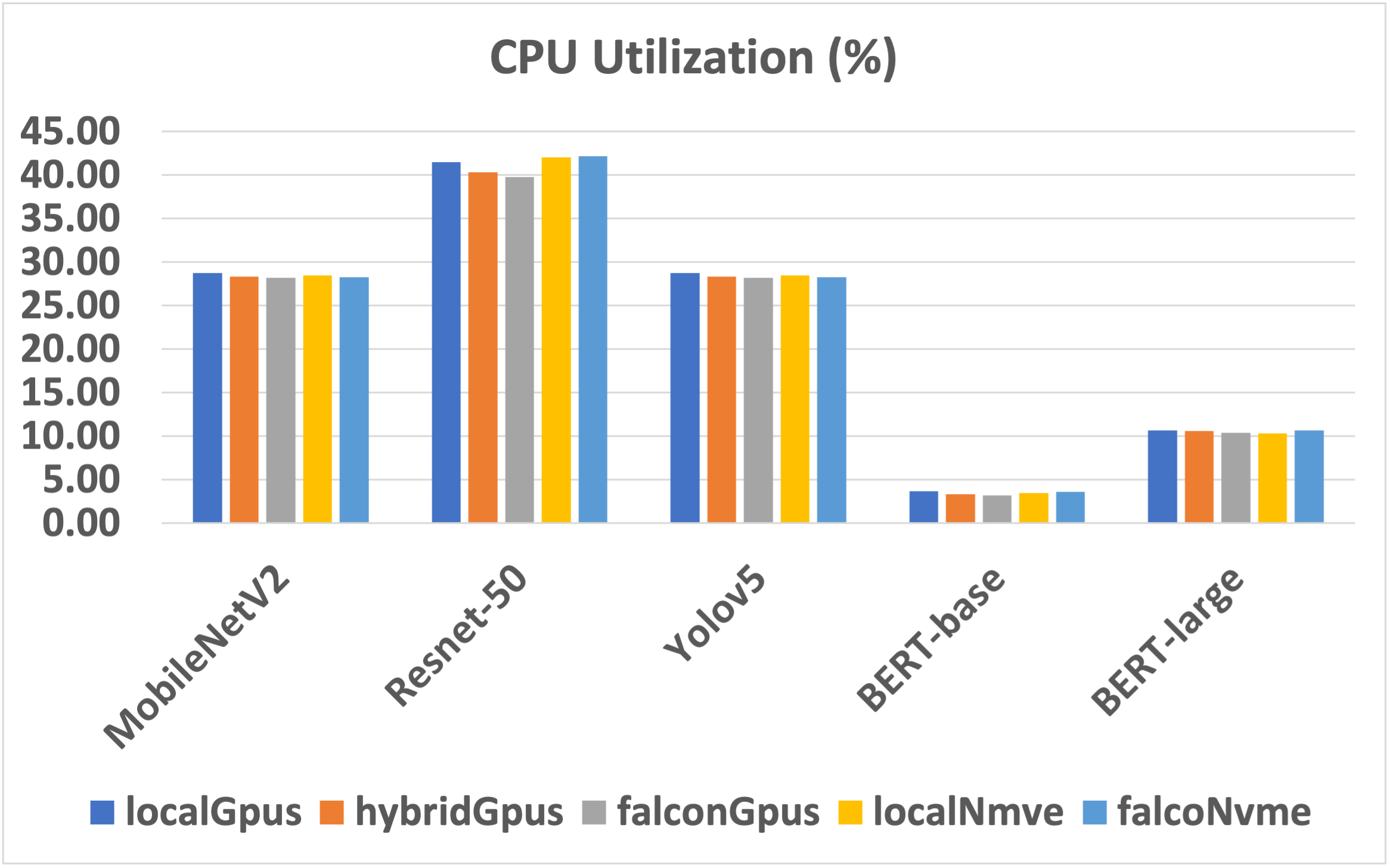}
    \caption{CPU Utilization of the DL Benchmarks on the Composable Configurations}
    \label{fig:cpu-util}
\end{figure}

\begin{figure}[ht!]
    \centering
    \includegraphics[width=8cm]{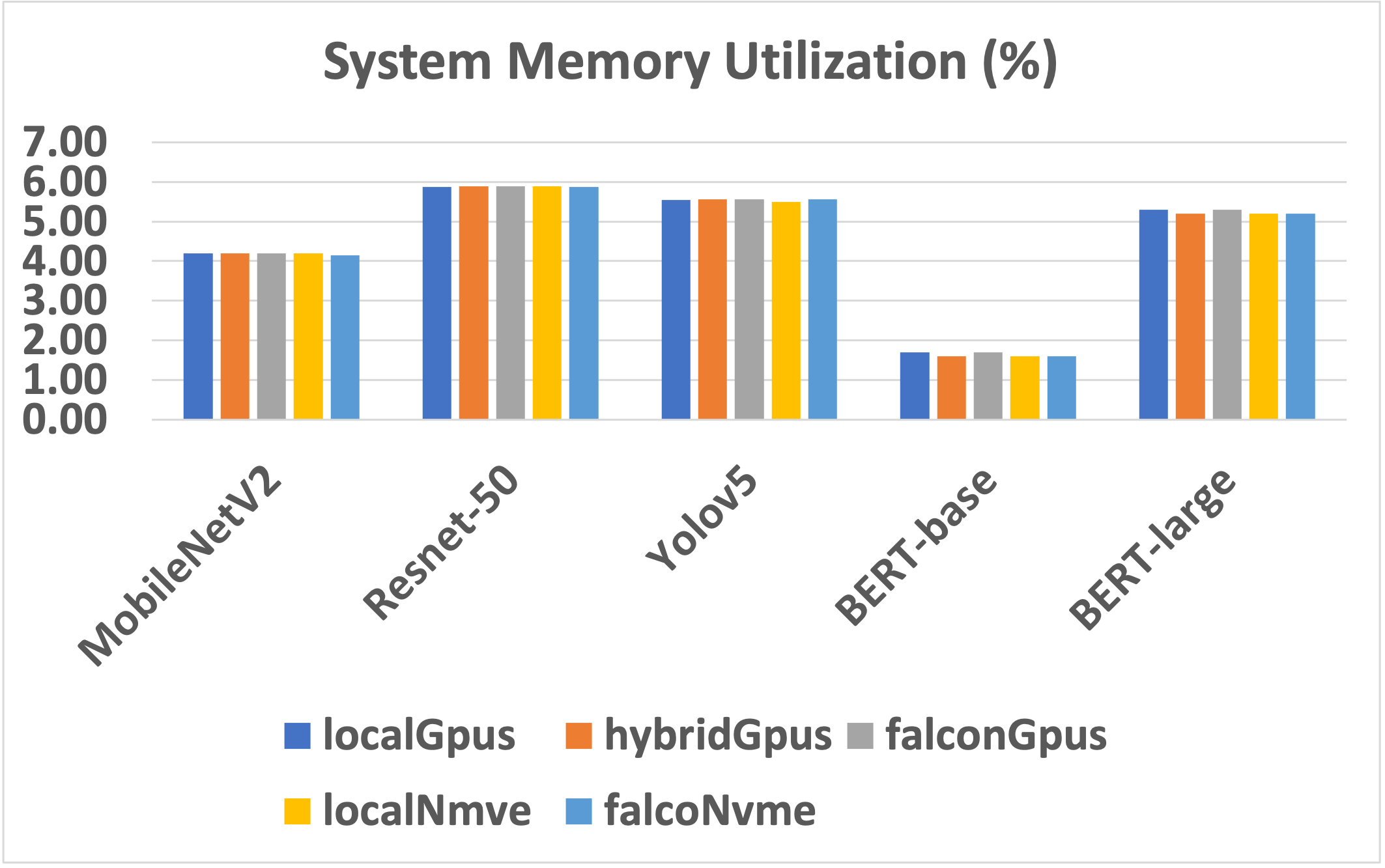}
    \caption{System Memory Utilization of the DL Benchmarks on the Composable Configurations}
    \label{fig:mem-util}
\end{figure}

To gain more insights into how the benchmarks are exercising the different components of the system, we have measured various system-level metrics such as GPU utilization, GPU memory utilization, CPU and system memory utilization and Falcon PCI-e traffic. Figure~\ref{fig:gpu-perf} shows the GPU utilization, GPU memory utilization and percentage of time the model spent accessing GPU memory. We notice that across all configurations, the behavior of GPU access patterns of the DL benchmarks is similar. We notice that the GPU utilization is slightly higher in the case of configurations that involve the Falcon GPUs, while the GPU memory access time is lower, especially in the case of the BERT benchmarks. All benchmarks are keeping GPUs busy most of the time with GPU utilization higher than 80\%. BERT models use the Transformer architecture which requires massive amounts of GPU memory.
Figures~\ref{fig:cpu-util} and~\ref{fig:mem-util} show that the benchmarks are not stressing the CPU cores or the system memory. The vision benchmarks are GPU compute bound while the NLP benchmarks are both GPU compute and GPU memory bound in our system. Vision benchmarks exercise the host CPUs more than NLP benchmarks and this is attributed to the data pre-processing that these benchmarks have to do on CPUs and that is not currently accelerated by GPUs (e.g., image random cropping, resizing, transposing, normalizing, etc.). 

\subsubsection{Falcon-attached Storage}
We also examine the impact of attaching an NVMe storage device to the Falcon switch on the training performance. We compare three configurations: falconNVMe, localGPUs, and localNVMe. All these configurations use locally attached GPUs to the host. The difference is in the storage subsystem used. We observe in Figure~\ref{fig:nvme} that attaching NVMe storage provides additional acceleration to the training for large models such as BERT and Yolo as it improves the data loading speed. The overhead of PCI-e switching through the falcon is small in this case. 
\begin{figure}
    \centering
    \includegraphics[width=8cm]{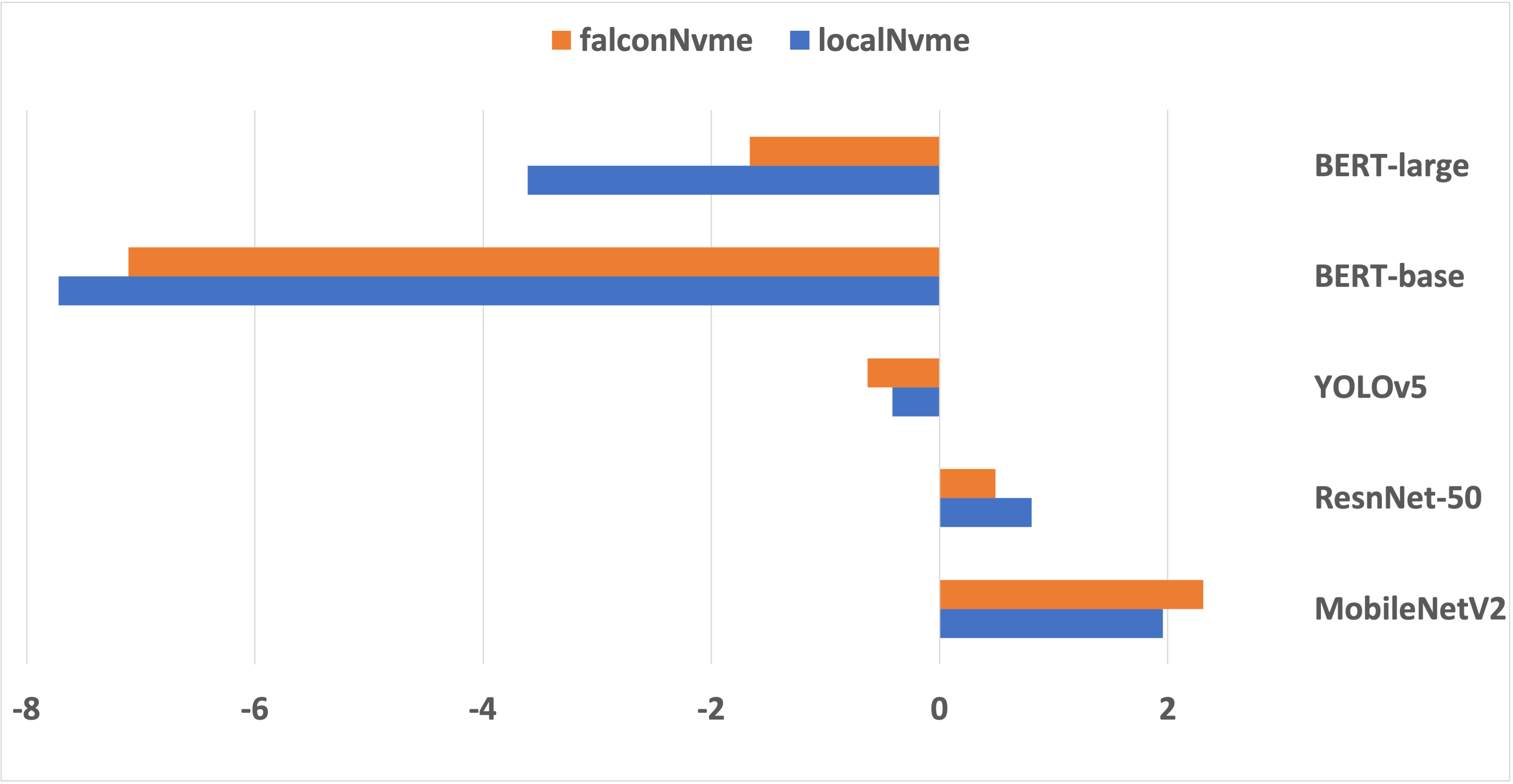}
    \caption{Percentage Change of Training Time from the Local GPUs Configuration}
    \label{fig:nvme}
\end{figure}

\subsubsection{Software-level Optimizations}
%So far we evaluated various DL benchmarks on different topologies of our composable infrastructure to validate the have looked at different system-level typologies, 
In this section, we examine the impact of existing deep learning software-level optimizations. All the experiments we have done so far use NVIDIA's FP16 mixed-precision training and PyTorch's highly optimized Distributed Data Parallel library (DDP). We compare the performance of BERT-large benchmark using PyTorch one node Data Parallel, Distributed Data Parallel, mixed precision training with FP16~\cite{micikevicius2018mixed}, and sharded training~\cite{rajbhandari2020zero}. 
Mixed-precision training provides significant speedup as it allows reducing the memory footprint during training while retaining the same model accuracy. This implies less communication overhead for synchronizing the models replicas among the GPUS. 

In addition to DDP, PyTorch also provides the Data Parallel (DP) mechanism when training on one node with multiple GPUs. In DP, one GPU maintains a master copy of the model parameters and broadcasts the parameters to all other GPUs at every iteration. The parameters are then all reduced back to the master GPU, which updates the model parameters. DDP on the other hand uses multiprocessing and distributes the burden model parameter synchronization among more than one process. DP is easier to debug; however, it often causes poor GPU utilization especially for large models. 

The sharded optimization further optimizes memory and is especially useful for large models. It tries to eliminate memory redundancies hence retaining low communication volume.
As shown in Figure~\label{}, mixed precision provides significant speedup:more than 50\% in all cases and more that 70\% in the case of Falcon-attached GPUs. DDP also provides additional speedup especially in the case of locally-attached GPUs (more than 80\%),  Using sharded optimization, we were able to increase the batch size from 6 to 10 which helped provide additional training time speedup.
\begin{figure}
    \centering
    \includegraphics[width=8cm]{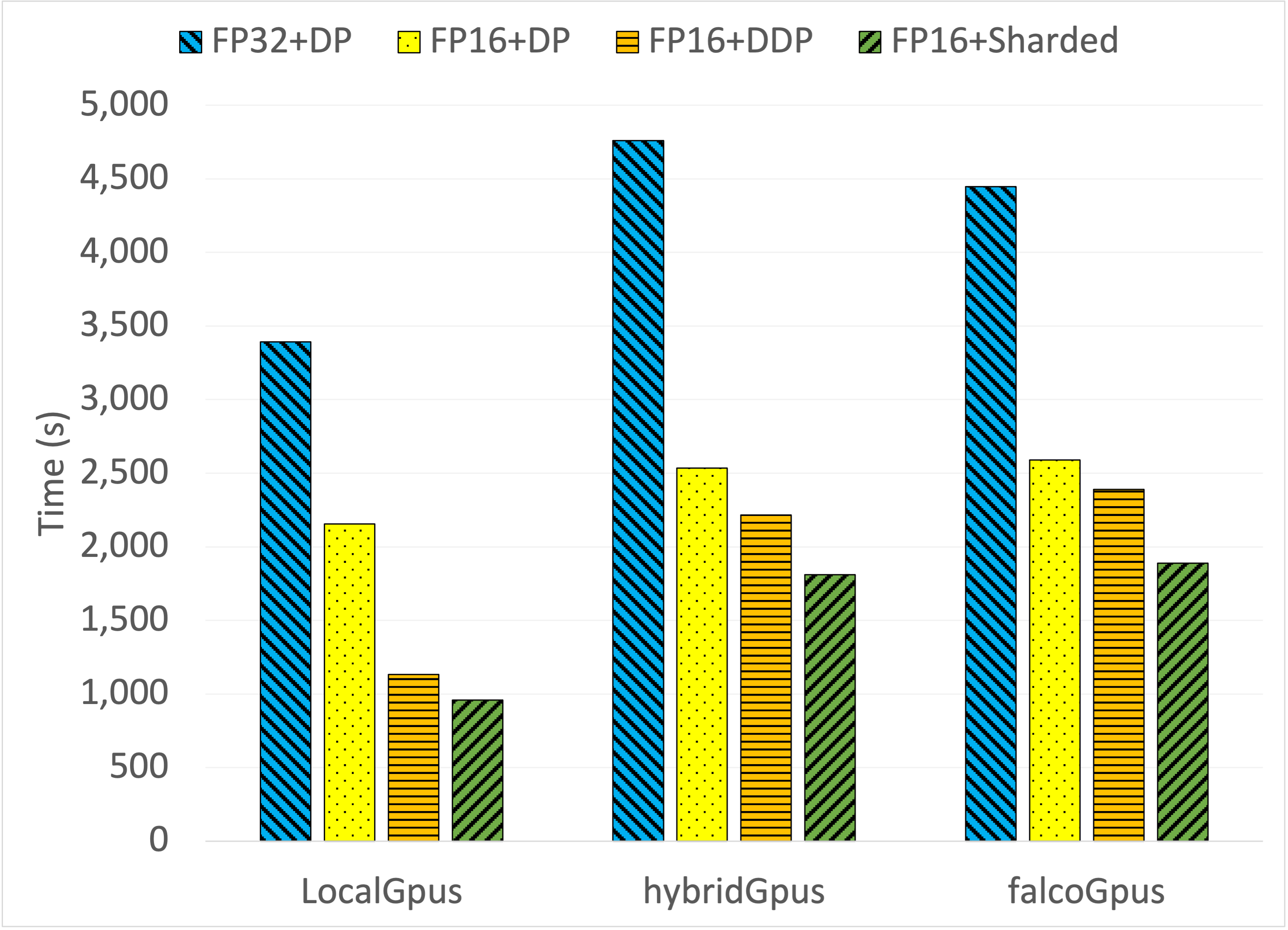}
    \caption{The Impact of Various Software-level DL Optimization on BERT-large Fine-tuning using the SQuaD dataset.}
    \label{fig:nvme}
\end{figure}
\section{Conclusions}
Our work is focused on using the composable server as an experimentation platform for new hardware, combinations of hardware accelerators, storage, network cards, etc. This design provides flexibility to serve a variety of workloads and provides a dynamic co-design platform that allows experiments and measurements in a controlled manner.%This environment provides an excellent experimentation platform, since we do not need to replicate current system network and memory speeds, as is required for use of a composable server in a production environment.

This works describes the first deep learning characterization on a composable infrastructure. As observed in our performance evaluation, the PCI-e switching slows down DL training by less than 7\% in the case of vision models. However it adds more overhead as the models become much larger, as in the case of large NLP models. However, we believe that such overhead is still acceptable given the flexibility that this type of platform provides without introducing any errors in the application behavior. We show that additional software-level optimizations can be used to mitigate the PCI-e overhead. %The flexibility of the platform allows for rapid re-configuration and the opportunity to quickly run benchmarks with tunable options.  We found that software-level performance tuning of distributed learning benchmarks and DL frameworks have a large impact on performance which the composable environment is well-suited to explore.

Our future work will investigate incorporating other accelerators and system components into the composable systems and further system-level optimizations. This will allow for a richer set of experiments. We plan also evaluate other modes of the system, such as advanced mode and dynamic reconfiguration, and finally, build a system framework that can take the input of various configured runs, and recommend the optimal system level topology for AI and High Performance Computing (HPC) workloads.

\section*{Acknowledgements}
The authors would like to thank the following colleagues from IBM Research: I-Hsin Chung and Paul Crumley for their technical assistance with early design and experimentation on the composable server.  James Norris for set up and implementation of the infrastructure.  Xiao Sun for his assistance with developing the benchmark assets.  We also acknowledge support from the IBM  Research AI Hardware Center and the Center for Computational Innovation at Rensselaer Polytechnic Institute for computational resources on the AiMOS Supercomputer.

%%%%%%% -- PAPER CONTENT ENDS -- %%%%%%%%

%%%%%%%%% -- BIB STYLE AND FILE -- %%%%%%%%
\bibliographystyle{IEEEtran}
\bibliography{refs}
%%%%%%%%%%%%%%%%%%%%%%%%%%%%%%%%%%%%

\end{document}